# ResumeVis: A Visual Analytics System to Discover Semantic Information in Semi-structured Resume Data

Chen Zhang
Institute of Software, Chinese
Academy of Sciences, Beijing, China
008613811513670
zhangchen@iscas.ac.cn

Hao Wang
Institute of Software, Chinese
Academy of Sciences, Beijing, China
008615001098360
wanghao@iscas.ac.cn

Yingcai Wu
State Key Lab of CAD & CG, Zhejiang
University, Hangzhou, China
008613738047313
ycwu@zju.edu.cn

## ABSTRACT

Massive public resume data emerging on the WWW indicates individual-related characteristics in terms of profile and career experiences. Resume Analysis (RA) provides opportunities for many applications, such as talent seeking and evaluation. Existing RA studies based on statistical analyzing have primarily focused on talent recruitment by identifying explicit attributes. However, they failed to discover the implicit semantic information, i.e., individual career progress patterns and social-relations, which are vital to comprehensive understanding of career development. Besides, how to visualize them for better human cognition is also challenging. To tackle these issues, we propose a visual analytics system *ResumeVis* to mine and visualize resume data. Firstly, a text-mining based approach is presented to extract semantic information. Then, a set of visualizations are devised to represent the semantic information in multiple perspectives. By interactive exploration on *ResumeVis* performed by domain experts, the following tasks can be accomplished: to trace individual career evolving trajectory; to mine latent social-relations among individuals; and to hold the full picture of massive resumes' collective mobility. Case studies with over 2500 online officer resumes demonstrate the effectiveness of our system. We provide a demonstration video[1].

## 1. INTRODUCTION

Massive publicly accessible resumes have emerged on the Internet. They outline people's attributes, experiences and qualifications previously and currently [1]. Therefore, Resume Analysis (RA) is an important means for personnel assessment. For example, with resumes one can assess people's current qualifications or predict their behaviors in the future by analyzing past experiences [2]. In general, RA has many applications, such as talent seeking, recruitment and job-performance evaluating in company's Human Resources department, and cadre assessment and appointment in government's think-tank department.

[1] https://1drv.ms/v/s!AstTHdadq8_-gz6Tku2v10TwhGzi

Besides, resumes can also reveal implicit people's career advancement process and latent social relations. For example, a government-officer or scholar's resume shows how his or her title/rank advanced from lower level through higher level. The implicit social-relations can be detected by identifying intersections of career experiences, e.g., co-worker relationship. We believe that such semantic information hidden in semi-structured resume text is vital for in-depth RA tasks, such as semantic searching, talent recommendation, career prediction and social network analysis.

However, existing RA studies neglected to discover such information. Instead, they were statistical analysing-based and basically focused on identifying explicit attribute information from structured resumes to solve problems such as talent recruitment. Besides, how to design the appropriate visualizations for enhancing human cognition in order to perceive career patterns and mobility from a huge amount of resumes is also a challenge.

To address these challenges, we firstly define three semantic analytical tasks in multiple perspectives:

- How to trace individual career evolving trajectory and uncover common growth pattern? (**Individual perspective**)
- How to mine latent social-relations among individuals? (**Interpersonal perspective**)
- How to hold the full picture of massive resumes for visualizing collective mobility? (**Collective perspective**)

Then, we propose a visual analytics prototype called *ResumeVis* to mine and visualize resume data. Specifically, a text-mining based approach is presented to discover semantic information, such as personal career paths and interpersonal social-relations, from semi-structured resume text. Besides, three visualization modules are devised to represent the mined semantic information:

- *career trajectory chart*. A timeline-styled chart which represents the career progress "trajectory" of an individual.
- *interpersonal relationship graph*. An ego-network based graph which reveals relations between individuals.
- *organization-individual mobility map*. A map which employs "quadrant diagram" metaphor to indicate people mobility.

Based on that, *ResumeVis* provides an integrated visual analytics environment, where users can gain knowledge and insights through visual interaction and in-depth exploration. Finally, the previously defined three semantic analytical tasks can be accomplished.

To the best of our knowledge, visual analytics on semantic information in resume data has not been researched previously. Based on *ResumeVis*, the analysis of resume can be extended into the analysis of people, which may trigger the application of *ResumeVis* in other fields such as public opinion monitoring.

**Fig. 1. An example of officer resume.**

The rest of this paper is organized as follows: Section 2 describes the input data, design requirements and an overview of our system. Section 3 proposes a semantic information extraction approach. Section 4 presents the visual design of our system. Case studies are performed in Section 5. Section 6 introduces related work. The conclusion is summarized in Section 7.

## 2. SYSTEM OVERVIEW

### 2.1 Description of Input Data

Electronic resume can be classified into two forms: **structured (sr)** and **semi-structured (smr)**. *sr* is commonly table-based and exists in database. The data format of *sr* is well-organized and usually fixed, so *sr* can be easily stored and managed. However, it is hard to acquire sufficient *sr* for high-level semantic analysis because the available *sr* on the Internet is limited and most of *sr* as internal resources are private. On the other hand, *smr* is a type of text document and usually exists on the Internet. The data format of *smr* is diverse and inconsistent because of text forms (txt, word, pdf, etc.), so it is non-trivial to uniformly manage and analyze *smr*. However, *smr* as publicly accessible resources can be obtained easily, which provide great opportunities for semantic based intelligence analysis. The examples of intelligence analysis are semantic search, person classification, person profile, etc [8].

**We use *smr* crawled on the Internet as input**. *smr* can be instantiated to government-officer's resume, enterprise-stuff resume, executive resume, scholar resume or pop-star resume. Specifically, **we use government-officer's resume in our system**. However, the analysis approaches and visualizations are general; it is easy to make an extension to address other types of *smr*.

A resume usually contains two sections: *basic information* (**BI-section**) and *experience* (**E-section**) [35]. *BI-section* covers a person's name, gender, age, date of birth, etc. *E-section* summarizes a person's education and/or career experiences.

Fig. 1 shows an example, which translates a Chinese resume. Due to the difference between Chinese and English, we add semicolons so that the pauses in a sentence can be easily recognized. In Fig. 1, the first 3 sentences constitute *BI-section*. Sentence #2 gives time stamps of two milestones, and sentence #3 outlines the current career title. The rest sentences, i.e., paragraphs #2 to #7, constitute *E-section*. Each paragraph describes an experience record containing time span, title, organization and location. Another case is that all experience records are within one paragraph. We omit the education experiences which have the similar structure with career experiences, and we only focus on the analysis of career experiences in this paper. However, the approach is general and can be extended into the analysis of education experiences.

### 2.2 Design Requirements

Our system is under a three-year International Innovation Team project founded by K. C. Wong Education Foundation. This project aims at public opinion surveillance from online media. Five domain experts are incorporated to supervise this project, including a vice head of an institute who is in charge of this project, two government administrators whose duties are cadre management and training, an intelligence analyst, and a HR expert in a company. Regular research discussion meetings and unscheduled idea exchange meetings with these experts have held since the year 2014. During these communications, the experts have clarified their requirements and evaluated the prototypes developed for the project.

#### 2.2.1 Requirements in mining phase

**R1 Semantic elements extraction.** Firstly, to extract career-related information mentioned in Section 2.1. Secondly, to quantify the career rank of the resume's title at each career stage. Taking officer title as an example: President, vice President, Governor, vice Governor, Mayor, vice Mayor, County Head, vice County Head and civilian are quantified as rank 8 through rank 0.

**R2 Pattern discovery.** Firstly, to mine personal career progress patterns. A domain expert claims that an effective feature for pattern discovery may be the resume's progress rate at each career stage. Secondly, to discover interpersonal social patterns based on the social relations among resumes. The social relations may be explicit where the intersection between individuals' career experiences exists. The relations may also be implicit if individuals' career progress paths are similar. The interpersonal social patterns can be further used to construct a social network, upon which numerous network analysis tasks can be performed.

**R3 Resume validation.** Given an unknown resume with no person name, incomplete, outdated or even wrong information, to identify who this resume may belong to and validate the incorrect information in this resume. Applications of this requirement could be updating the outdated resumes, or revising the inaccurate resumes produced intentionally or by accident.

#### 2.2.2 Requirements in visualization phase

**R4 Representing the resume from overview to details.** When dealing with numerous resumes, users need a multi-level visualization so as to investigate career paths from overview to details and perceive career progress patterns comprehensively.

**R5 Facilitating visual comparisons between different resumes.** The visualization should facilitate the comparison between the resumes with various progress patterns and social patterns through metaphor representation, where the career path of a resume and the social relations between resumes can be intuitively portrayed (Figs. 5a & 5b).

**R6 Visualizing collective mobility from massive resumes.** The visualization should summarize the collective mobility across temporal dimension and organization dimension so that users can quickly hold the full picture of numerous resumes and gain insights easily. For example, the visualization should reveal the transfer process of people between various communities, such as government, grass roots, state-owned enterprises and non-profit organizations (Right part of Fig. 2).

**R7 Flexible data labeling to collect user feedback.** The pattern discovery model in **R2** depends on predefined categories and labeled data. In order to incorporate human knowledge and improve model performance, the visualization and corresponding interactions should be designed to support smoothly data labeling and model tunning (Fig. 6).

**R8 Easily browsing of raw data.** The raw data such as text in each resume is the strongest support for validating the semantic elements and patterns discovered by the system. The visualization should enable users to explore the raw data easily (Fig. 6).

### 2.3 System Workflow

Fig. 2 illustrates the system workflow. It has three phases:

The mining phase will be described in Section 3. It loads input data, extracts semantic elements and mines pattern information.

The visualization phase and interaction phase will be described in Section 4. The former contains three visualization modules supporting visual exploration and analytics of mined semantic information. The latter forms a loop between mining and visualization, where users are involved for perceiving, gaining insights and generating feedback, such as tuning parameters.

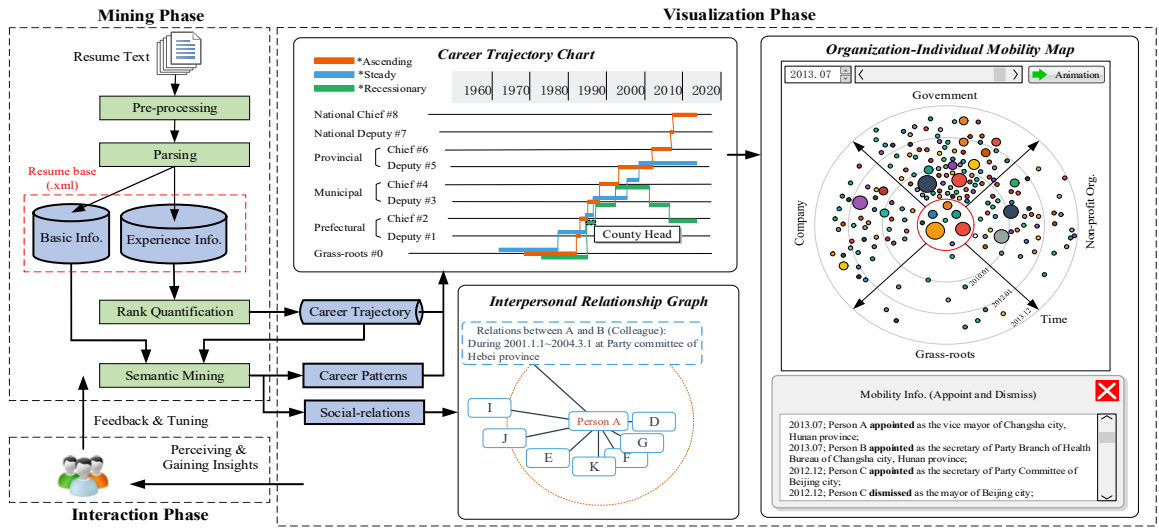

**Fig. 2. Overview of our system workflow.**

## 3. SEMANTIC EXTRACTION

In left part of Fig. 2, the approach contains four components: pre-processing, parsing, rank quantification and semantic mining.

### 3.1 Parsing

We propose a method to parse semi-structured resume text using Chinese Word Segmentation and NER techniques (**R1**). Fig. 3 shows an example of extraction result based on the resume in Fig. 1. In Fig. 3, **BI-section** and **E-section** are extracted and saved into structured resume base as .xml format (Fig. 3).

**BI-section** includes some tags such as /Name, /Gender and /Nation. We use an information extraction technique [14] to extract these tags.

**E-section** contains a list of experience records in a chronological order. Each record consists of some tags: /DateBegin, /DateEnd, /Location, /Org, and /Title. In the case that all experience records are within one paragraph of the original resume, we use /DateBegin and /DateEnd as separators to divide one paragraph into several paragraphs. We use the similar technique [14] to extract tags /DateBegin, /DateEnd, /Location, /Org and /Title. Specifically, we choose Chinese characters "*year*", "*month*" and "*day*" as the keywords to extract the time range of each experience record. We choose Chinese characters "*province*", "*city*", "*county*", "*town*", etc. to extract the location of each experience record. We choose Chinese characters "*division*", "*bureau*", "*department*", "*institute*", "*academy*", etc. to extract the organizations of each experience record. We use Chinese characters "*head*", "*secretary*", "*committee member*", etc. to extract the titles of each experience record. These keywords are collected from a large number of Chinese government websites.

In Fig. 3, we use the extracted semantic elements to form a resume base. The resume base is designed as .xml format with the tree structure of the experience section. On base of structured resume base, the subsequent data mining and visualization techniques can be easily performed. Due to space limitation we only show one experience record in Fig. 3 while the rest are omitted.

Finally, the semi-structured resume text has been converted into the structured resume base.

### 3.2 Rank Quantification

We quantify the career rank of each record, (<rank> in Fig. 3). We take administrative levels mentioned in Chinese Civil-Servants Law [36, 37] as the reference in which quantification rules are extracted. For example, the mayors of the ordinary cities should be quantified as rank 4 (**R1**). However, there are many exceptions. For example, the mayors of *Beijing*, *Shanghai*, *Tianjin* and *Chongqing* should be quantified as rank 6, because these four cities are municipalities which directly under the central government and have higher rank than ordinary cities.

Eventually, we extract a career trajectory sequence based on these rules and exceptions. Table 1 shows an example of the career trajectory sequence extracted from the resume base in Fig. 3.

Please note each section in the career trajectory sequence, including rank value, can be modified by users in UI shown in Fig. 6. This is essential when the information is incorrect due to the complexity of Chinese administration and numerous special cases.

### 3.3 Semantic Mining

After obtained the resume base with quantified career rank, we perform various mining methods to discover semantic patterns (**R2**).

```xml
<?xml version="1.0" encoding="UTF-8"?>
<resume_base version="1.0">
  <resume category="1"> //ID of this resume
    <basic_info_section>
      <name>Jim</name> <nation>Han</nation> <birth_place>Changsha city,
Hunan province</birth_place>
      <age>39</age> <gender>Male</gender>
      <date category="birth"> <year>1975</year> <month>5</month>
<day>1</day> </date>
      <date category="party"> <year>1991</year> <month>12</month>
<day>1</day> </date>
      <date category="work"> <year>1990</year> <month>1</month>
<day>1</day> </date>
    </basic_info_section>
    <experience_section>
      <experience_record>
        <date category="begin"> <year>1989</year> <month>1</month>
<day>1</day> </date>
        <date category="end"> <year>1992</year> <month>1</month>
<day>1</day> </date>
        <location> <province>Hunan</province> <city>Changsha</city> </location>
        <organization>
          <organization_name>Party Branch of Health Bureau</name>
          <title_array>
            <title> <title_name>Secretary</post_name><rank>3</rank> </title>
          </title_array>
        </organization>
      </organization_array>
      </experience_record>
      //The rest of experience records are omitted
    </experience_section>
  </resume>
</resume_base>
```

**Fig. 3. The resume base built based on raw resume in Fig. 1.**

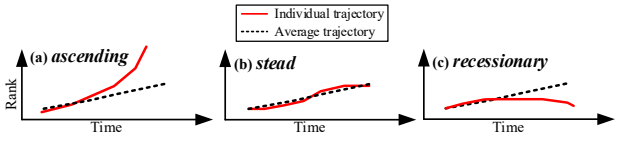

**Fig. 4. Three career progress patterns.**

**Table 1. The career trajectory sequence**

| /DateBegin | /DateEnd | /Location | /Org | /Title | /Rank |
|---|---|---|---|---|---|
| 1989.1.1 | 1992.1.1 | Changsha, Hunan | Party Branch of Health Bureau | Secretary | 0 |
| 1992.1.1 | 1995.1.1 | idem | County Party Committee | County head | 2 |
| 1995.1.1 | 1998.1.1 | idem | Municipal Party Committee | Vice mayor | 3 |
| 1998.1.1 | 2002.1.1 | idem | Municipal Party Committee | Mayor | 4 |
| 2002.1.1 | 2010.1.1 | idem | Provincial Party Committee | Vice governor | 5 |
| 2010.1.1 | 2015.12.11 | idem | Provincial Party Committee | Governor | 6 |

### 3.3.1 Discovering individual career patterns

We employ Naïve Bayes [18] to discover individual patterns. The resumes are classified into various categories based on their experiences information; each category is considered as a career progress pattern. The unlabeled resume can be automatically classified by the model trained on labeled resumes.

**1. Definition of career progress patterns**

Taking the mean of all resumes' trajectories as baseline, the domain experts defined three types of career progress patterns to describe resumes with various career trajectories: *ascending*, *steady* and *recessionary*. In Fig. 4, *ascending* represents the resume that the rank growth rate (red line) is observably larger than the average rate (black dot line) across the whole time dimension; *steady* represents the resume that the rank growth rate is roughly equal to the average rate across the whole time dimension; *recessionary* represents the resume that the rank growth rate is observably smaller than the average rate across the whole time dimension. Specifically, the rank of *recessionary* can even decrease in real cases. The trajectory lines in Fig. 4 depict the rank growth rate along the time axis, which are calculated based on the time span of each rank value in Table 1.

Please note these three patterns are specified by the domain experts based on their knowledge and observation. There are other more complicated patterns, such as the combination of the three patterns in different career stages. Here we choose three basic patterns to demonstrate our system's functionality. Also, the labeling process, in which each resume in the training set is assigned as one pattern, is conducted by the domain experts.

**2. Selection of career progress features**

We take the time span on each rank value as the feature to describe the career progress pattern. It can be formalized as (1).

$$\{r_1 = t_1/T, r_2 = t_2/T, ..., r_n = t_n/T\} \quad (1)$$

where $n = 9$, which denotes the number of rank value. $T = \sum_{i=1}^{n} t_i$, which denotes the time span across whole career experiences. For example, the feature of the sequence in Table 1 is formalized as $\{0.11, 0, 0.11, 0.11, 0.15, 0.3, 0.22, 0, 0\}$, where $T = 27$.

**3. Training classifier using labeled resumes**

Let $x = \{r_1, r_2, ..., r_n\}$ as a resume's feature set; $C = \{y_1, y_2, y_3\}$ as the pattern set, where $y_1$, $y_2$ and $y_3$ separately denote *ascending*, *steady*, and *recessionary*.

Firstly, we calculate the conditional probabilities by using (2).

$$P(r_i \mid y_j) = g(r_i, \eta_{y_j}, \sigma_{y_j}) = \frac{1}{\sqrt{2\pi}\sigma_{y_j}} e^{-\frac{(r_i - \eta_{y_j})^2}{2\sigma_{y_j}^2}} \quad (2)$$

where $\eta_{y_j}$ and $\sigma_{y_j}$ separately denote the mean and standard deviation of $r_i$ on the pattern $y_j$. Because $r_i$ is real number, we assume that $r_i$ follows a Gaussian distribution.

Secondly, for each unlabeled resume $x$, we calculate the posterior probabilities of each pattern $y_j$ by using (3).

$$P(y_j \mid x) = \frac{P(x \mid y_j)P(y_j)}{P(x)} = \frac{P(y_j)}{P(x)} \prod_{i=1}^{n} P(r_i \mid y_j) \propto P(y_j) \prod_{i=1}^{n} P(r_i \mid y_j) \quad (3)$$

Thirdly, we classify $x$ as $y_k$, where $k = \arg\max_k \left[ P(y_j \mid x) \right]$.

### 3.3.2 Mining interpersonal social-relations (**R2**)

**1. Explicit relation mining**

Let $\mathbf{M} = \{M_1, M_2, ..., M_n\}$ as the resume base containing $n$ resumes. Firstly, we construct a basket dataset where the organization is considered as the "basket". By scanning $\mathbf{M}$, a list of organizations $\mathbf{O} = \{O_1, O_2, ..., O_m\}$ is obtained. For each organization $O_i$, we add a line into the basket dataset; the elements in this line are the resumes which belong to, or belonged to $O_i$.

Secondly, we employ Apriori [24] to mine frequent resume set.

Thirdly, we propose a resume matching algorithm to measure the matching degree between each resume pair in the frequent resume set. The basic principle of the algorithm is that more experience intersection produce higher matching degree; the matching degree represents the relation value between the resume pair. The value range of matching degree is from 0 (totally mismatching) through 1 (exactly matching, which means two resumes are the same).

Given an unknown resume $M_u$ with no person name, incomplete or even incorrect information, we can identify a resume $M_c$ from the resume base $\mathbf{M}$ so that the matching degree between $M_u$ and $M_c$ reaches a peak value. It indicates that $M_u$ most potentially belongs to $M_c$. Besides, the possible incomplete or wrong information of $M_u$ can be validated based on the experience intersection between $M_u$ and $M_c$ (**R3**). Please refer to *resume validation view* in Fig. 7 for details.

**2. Implicit relation mining**

We calculate the cosine similarity between resumes to measure the implicit relations. The features used are the career progress features mentioned in Section 3.3.1. Higher similarity indicates that the career trajectories of resumes are more similar.

Given a resume $M$, top $K$ resumes can be extracted from the resume base $\mathbf{M}$ based on explicit relation and implicit relation mining. These similar resumes will be visualized in *interpersonal relationship graph* in Fig. 5(b).

## 4. VISUALIZATION DESIGN

To satisfy the requirements in visualization phase (**R4~R8**), we provide our system with three visualization modules.

### 4.1 Career Trajectory Chart

We present a timeline-based chart (Fig. 5a) to visually encode the career progress process of an individual. This chart as a variant of classical line chart converts the abstract individual career evolving information into an intuitive representation. Specifically, the temporal variation of individual's rank, such as official title, and the promotion/demotion timestamp are visualized. In addition, this chart can assist users in understanding various career progress patterns intuitively.

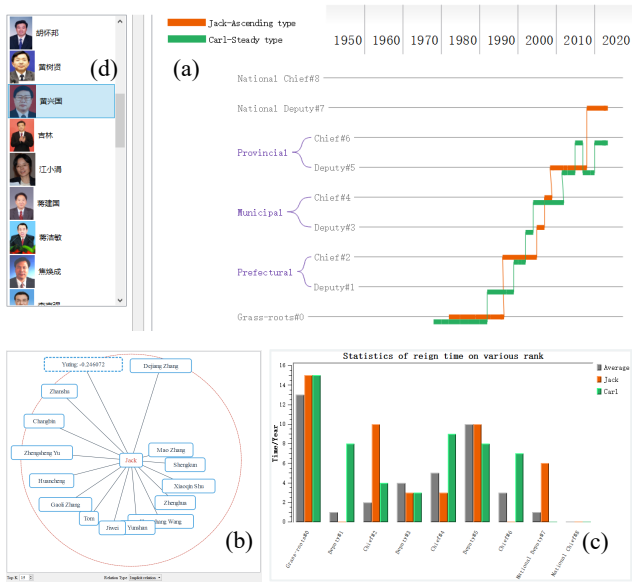

**Fig. 5. Main window of *ResumeVis*. (a) *career trajectory chart*. (b) *interpersonal relationship graph*. (c) *statistical histogram*. (d) *object manage panel*.**

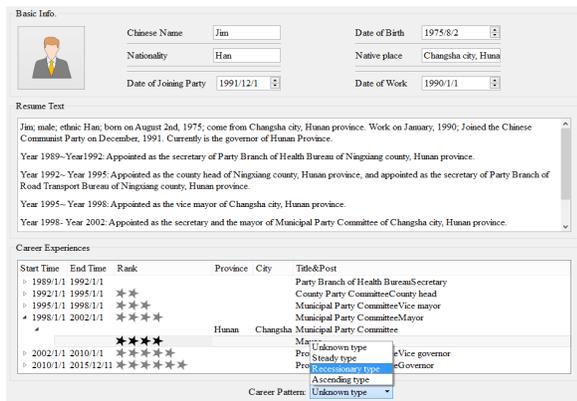

**Fig. 6. Basic information and setting window.**

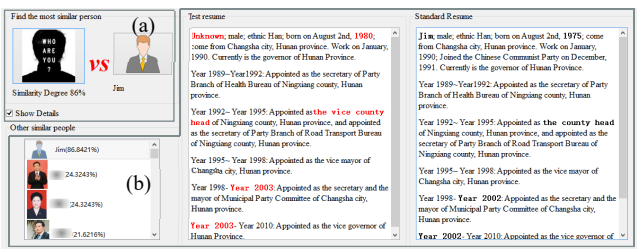

**Fig. 7. Resume validation view. (a) The collapsed view; (b) The expanded view.**

In Fig. 5a, *x*-axis encodes the temporal dimension. There are two modes, "*age*" and "*year*", which are toggled alternately by pressing the corresponding buttons. In addition, *y*-axis encodes the dimension "rank", i.e., quantified rank value of the career trajectory sequence. The ordinal scale is from rank 0 to rank 8. In this paper, these rank values are mapped into levels of the administrative titles from Civilian (0) through President (8).

Then, the resume's career trajectory can be visualized as a set of color-filled horizontal rectangles, each of which denotes an experience record. These rectangles are assembled head-to-tail by some vertical lines. Mouse hovering over a rectangle reveals detailed experience information via a floating tooltip window. To enhance dynamic evolving effect when rendering the trajectory, the animated transition is adopted so that the rectangles emerge one after another in a chronological order. Besides, the animated transition also occurs to preserve mental model when switching between mode *age* and mode *year*. Legends on this chart shows the names and the career progress patterns of corresponding resumes. Eventually, a "ladder" representation is constructed. To avoid overlapping among different trajectories, we add a tiny *y*-axis offset on trajectories for easily distinguishing them.

In Fig. 5c, we employ *statistical histogram* to get an overview of total resumes' trajectories. Besides, we can compare the statistics of trajectories between different resumes where the average value of all resumes (denoted as gray bars) is depicted as a baseline.

In Fig. 6, we use "Resume Text" view to display the raw resume as a baseline for easily browsing of raw data (**R4**).

Based on that, users can drag the icons of individuals on *object manage panel* (Fig. 5d) into the chart (Fig. 5a) to perform visual comparison, where different resumes are classified with various colors (**R5**).

Users can click the icon on *object manage panel* to popup a window to depict the basic information of the resume, as shown in Fig. 6. This window contains three sections: Basic Info., Resume Text and Career Experiences. In "Basic Info.", users can review and modify the basic information, such as uploading ID photo or editing name. In "Resume Text", users can review raw resume text. In "Career Experiences" section, user can modify the semantic elements of career experiences (**R8**). For example, rank values are indicated as stars; users can adjust the number of stars by clicking the stars when rank quantification is incorrect. After the rank values are modified, *career trajectory chart* with new rank values is updated smoothly.

Users can label the pattern type by clicking "Career pattern" combo box at the bottom of Fig. 6. Users can also modify the current pattern type predicted by the classifier in Section 3.3.2 (**R7**). This label/annotation function not only facilitates expressing the insight of users but supports the classification learning process.

### 4.2 Interpersonal Relationship Graph

We present an ego-network based spiral graph, in Fig. 5b, to reveal interpersonal relations between a focus person, denoted as a central node, and other related people, denoted as surrounding nodes. As a complement of *career trajectory chart*, this graph can enable users to understand the "context" information of the focus person.

When users right-click the icon of a resume $M$, the graph is constructed in which $M$ is visualized as the central point and $\mathbf{M}_k$ are shown as the $K$ surrounding nodes. The node is visualized as a rounded rectangle with the name of the individual on it. Please note $\mathbf{M}_k$ are calculated in Section 3.3.2. The distance between the central node and a surrounding node encodes the relation value between the focus person and a related person. That is, shorted edge means closed relations.

The parameter $K$ and relation type can be set via tool buttons at the bottom of the graph. The mouse hovering over a surrounding node shows detailed relation information. When the explicit relations are chosen, the relation value is shown as in Fig. 9a; otherwise, the experience intersection is shown as in Fig. 9b. Users can drag the surrounding node to adjust the position of the node along the reference circle for optimized layout. When dragging a node, an auxiliary red circle emerges as a baseline to facilitate the comparison of relation value among other nodes (see Fig. 5b).

When users load an unknown resume $M_u$ with no person name, incomplete, outdated or even wrong information, we propose *resume validation view* shown in Fig. 7 to identify who $M_u$ may belong to and validate the incorrect information in $M_u$ (**R3**). In Fig. 7a, the most similar person of $M_u$ is calculated as Jim and the similarity degree, i.e., matching degree in Section 3.3.2, between $M_u$ and Jim is 86%, which means 86% of Jim's career experience are matched with $M_u$. Users may click "Show Details" button to investigate the detailed information in the expanded view. In Fig. 7b, other similar people as candidates are listed in a descending order according to the similarity degree. After clicking an icon in the list which denotes a candidate $M_s$ is selected, $M_u$ is shown in "Test Resume" section and $M_s$ is shown in "Standard Resume" section. The mismatch between two resumes' text is highlighted as red bold font.

### 4.3 Organization-Individual Mobility Map

We propose a map for gaining insights on collective mobility through visualizing massive resumes. This map employs the metaphor "quadrant diagram" (right part of Fig. 2) to represent the collective mobility across temporal and organization dimension (**R6**). Here organizations are classified as four categories by considering the suggestions and comments of several domain experts. They are government, grass roots, state-owned enterprises (or companies) and non-profit organizations. Resumes belonging to same type of organization form a community. Therefore, resumes are divided as five communities: four basic communities and one compound community where individuals belong to multiple types of organizations.

In right part of Fig. 2, each node denotes an individual, of which the size encodes the rank value. The map is divided as five regions to encode five communities: four sectors denote four basic communities and the circle region in the center denotes the compound community. Four radials as the boundaries of four basic communities encode the dimension "time", which increases from center to around. Therefore, the location of the node encodes the type of organization to which the individual belongs at a specific time. As time evolves, the size and the location of the node vary. The transition process is visualized with the form of animation so as to reveal the collective mobility. A force-directed layout method [38] is applied to avoid node overlapping while keeping the relative positions of the nodes as much as possible. The detailed mobility information, with two forms of appointment and dismissing, is shown in "Mobility Info." window (bottom-right corner of Fig. 2).

Users can zoom-in, zoom-out and drag the map for exploration using various resolution. Users can select a node to investigate the detailed information regarding the individual and the organization(s) he/she belongs to (Figs. 10b and 10c). Users can also select a community to gain the information regarding the individuals belong to the selected community (Fig. 10d). When selecting a community, the nodes in this community are highlighted and the other nodes are hyalinized so that users can focus on interesting area. When users select a node from the compound community, several auxiliary lines emerge to indicate which communities this node belongs to (Fig. 10e).

In top-right corner of Fig. 2, users can press "Animation" button to play the transition process over time for perceiving the collective mobility during the whole time span. Besides, users can slide the slider so as to discover the status at a specific time stamp.

## 5. CASE STUDY

We evaluated the system through three case studies. In each case study some real-world tasks were designed by three of our domain experts. The other two experts were then required to use the system for solving the tasks. Before the evaluation the experts spent two hours using the system so that they were familiar with the system. Feedbacks from the experts were received after the evaluation. The dataset contains 2640 online resumes collected from websites of Chinese governments at various levels.

### 5.1 Individual Career Progress Analysis

**T1**: Summarizing the general rule of career development.
**T2**: Comparing various career paths of different resumes.
**T3**: Searching target officer based on semantic information.

These tasks are vital for government to select and promote potential officer. It is believed that selecting proper officer for further promotion is non-trivial because many aspects of the individual need to be considered, such as the career growth rate at various stage. However, there is no quantitative standard for assessing these aspects. Currently, this task is performed manually based on human knowledge and experience. When facing massive resumes, evaluating each resume merely by reading resume text is rather tedious and time-consuming.

In Fig. 8, after experts imported the resumes into our system, the system provided *statistical histogram* to indicate the statistical pattern of career progress. Fig. 8a shows the statistical results of total resumes, i.e., the average value of all resumes on each rank's time span. Figs. 8b, 8c and 8d separately denote the statistical results of the individuals named "*Jack*", "*Tom*" and "*Karl*", whose resumes were automatically classified as *ascending*, *steady* and *recessionary*, respectively. From this view experts can gain the following insights (**T1**):

1. There are two "bottleneck" period for Chinese officers during their early career. The first is rank#0. They usually spend 13 years at grass-roots organizations before they get the first promotion to rank#1, such as vice county head. The second is rank#5. They usually spend 10 years at rank#5, such as vice governor, before they get the promotion to rank#6, such as governor.

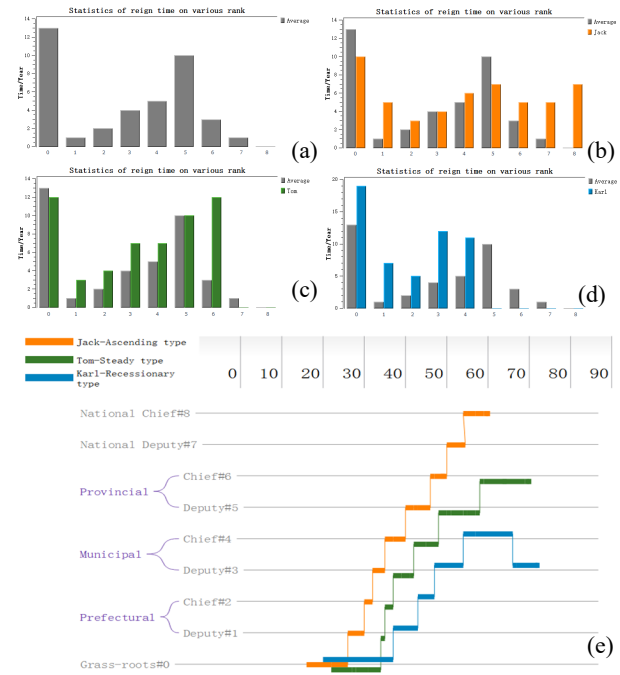

**Fig. 8.** The *statistical histogram* views (**a, b, c and d**) and *career trajectory chart* (**e**) of individuals' career trajectories.

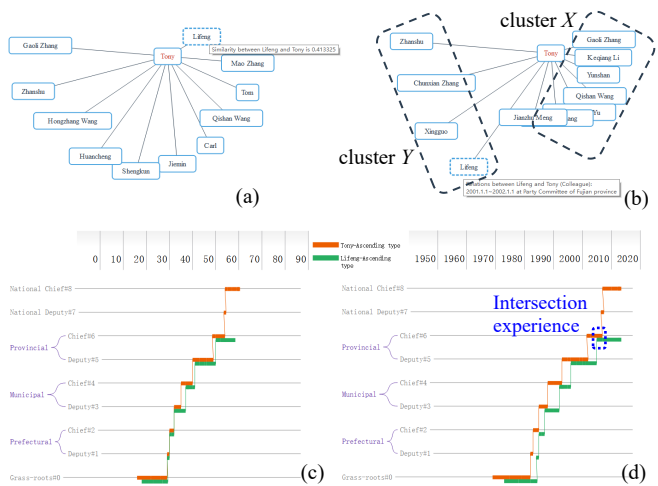

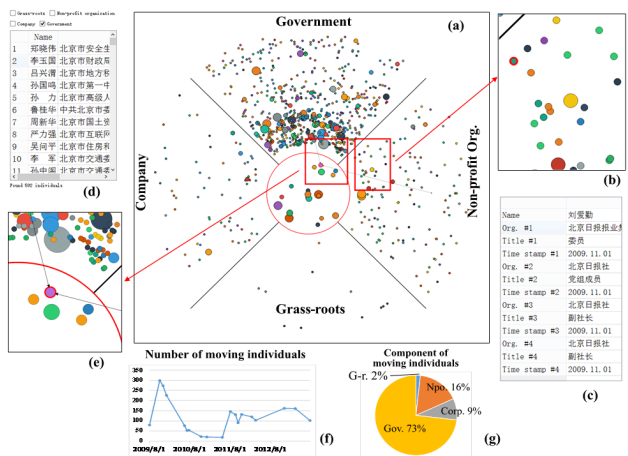

**Fig. 9.** *Interpersonal relationship graph*s (a: implicit relation; b: explicit relation) and corresponding *career trajectory chart*s (c: time mode "*age*"; d: time mode "*age*").

2. Officers with career progress pattern *ascending*, such as *Jack*, get faster promotion at the two "bottleneck" period than the average. Officers with pattern *steady*, such as *Tom*, have approximate promotion speed with the average at their early career. Officers with pattern *recessionary*, such as *Karl*, have slower promotion speed than the average at their early career.

Fig. 8e shows *career trajectory chart* with time mode "*age*" visualizing the career progress trajectories of *Jack*, *Tom* and *Karl*. From this chart experts easily perceived visual insights mentioned above by comparing career trajectories with various patterns from individual perspective (**T2**).

To perform **T3**, experts quickly skimmed the resumes with career progress pattern *ascending* (about 200 resumes) by visualizing them in *career trajectory chart*. Among these resumes experts then filtered 39 resumes as candidates of potential officers which have "sharp" trajectories. At last, experts selected 10 resumes as potential officers by investigating and examining the detailed semantic information of each resume shown in Fig. 6. The whole process took nearly 2 hours. In contrast, without our system experts may spend several days to read over 2600 resumes for selecting target resumes.

### 5.2 Interpersonal Social-relations Analysis

**T4**: Finding the similar or relevant resumes (individuals) based on the focus resume (individual).

**T5**: Validating unknown/ unfaithful resumes.

Please note **T4** is important for government to discover the interpersonal relationship and even social network in Chinese government. This task may facilitate anti-corruption.

When experts investigated *interpersonal relationship graph* of the focus individual named "*Tony*" shown in Figs. 9a and 9b, they may perceive the following phenomena (**T4**).

1. Through interaction with UI, e.g., switching the *Relation Type* and changing the value of *K*, experts observed that surrounding points can be classified as two clusters: clusters *X* and *Y*. By investigating the detailed information regarding the relations between the central point and clusters, experts realized that the duration of career intersection between the central point and *X* is significantly larger than that of *Y*. Besides, the intersection type, i.e., the organization type, of *X* is "*Company*" while the intersection type of *Y* is "*Non-profit Org*".

2. The individual *Lifeng* has close relations with *Tony*. Specifically, they have similar career progress trajectories in Figs. 9c and 9d. Another evidence in Fig. 9a shows the implicit relations, i.e., the similarity, between *Lifeng* and *Tony* reaches a peak value. In addition, the explicit relations in Fig. 9b shows that they were colleagues during year 2001. Based on that, experts can infer that the promotion of *Lifeng* may benefit from the intersection experiences with *Tony* (blue mark in Fig. 12(d)). Experts can even predict *Lifeng* will get further promotion soon based on such trajectories similarity and relations.

Experts can interact with the UI in Fig. 7 for performing **T5**. We collected 50 resumes from the officer base of Xinhua Net [3]. Two participants (not experts) were requested to delete 10%~30% of text content (including "name" section) from first 25 resumes for producing "unknown resumes", and modified 15% of text content from second 25 resumes for producing "unfaithful resumes". Such operations were performed by using our system as shown in Fig. 6. Based on these produced resumes, experts successfully identified 21 unknown resumes and 19 unfaithful resumes, as well as their missing or wrong information being corrected.

### 5.3 Collective Mobility Analysis

**T6**: Discovering career evolving patterns across temporal dimension.

**T7**: Discovering career transfer patterns across organization dimension.

Solving **T6** and **T7** enables decision makers to be aware of career situation of officers in a macroscopic perspective and thus make reasonable strategies accordingly, such as guiding and controlling the process of officer training, appointing and dismissing. Through animation and interaction with *organization-individual mobility map* in Fig. 10a, experts can gain two valuable insights.

First, there is a "booming" of officer emergence during 2009/11/01 and 2010/01/01 (The dense area of government section in Fig 10a). Fig. 10f proves this "booming" phenomenon, which shows the emergence trend during the whole time range. From this phenomenon domain experts inferred that the reason behind may be the fourth plenary session of the 17th central committee of the communist party of China held in Sep. 2009 (**T6**).

Second, the number of resumes in government significantly outweigh the number of resumes in other organizations. Specifically, the number of grass-roots' resumes is the least. Fig.

**Fig. 10.** (a) *organization-individual mobility map*. A selected node (b) and its information (c). (d) The information of the selected community (Government). (e) A selected node belonging to multiple communities. (f) and (g) are maps created by Excel based on the statistics of resumes.

10g can prove that, which shows the proportion among various organizations. This phenomenon demonstrates the fact that currently being a civil servant in the government is the best career choice for Chinese job seekers (**T7**).

Besides, experts can conduct in-depth exploration by ample interactive tools. For example, when experts select a node (Fig. 10b), the information regarding this node is shown in Fig. 10c. Experts can also select a community so that the nodes in this community are highlighted and the information of this community is shown in Fig. 10d. Fig. 10e indicates that experts can check the nodes belonging to multiple communities.

### 5.4 User Feedback

We collected the feedback from the two experts. In Table 2, the experts were required to rank a score on functionality and usability of our system for each task. The scores could be 5 (excellent), 4(good), 3 (average), 2 (not good) or 1 (poor).

Then, the experts gave their comments about user experience. Both experts were very impressed by the system, particularly by the revealed semantic information which can well support their decision making. They suggested that representing resumes in multiple perspectives is "comprehensive", and "valuable for human cognition and perceiving insights". In particular, the first expert highlighted the novelty of *career trajectory chart* for outlining career path. He said "Never have I seen such a visualization that can intuitively represent the career progress trajectory from the resume data". He was very impressed by investigating the resume data from overview to details. He believed that the metaphor employed in the system can definitely improve the efficiency of RA. He also mentioned that "the idea of representing social-relations between different resumes as a spiral graph is excellent for understanding social patterns among individuals". The second expert particularly liked *organization-individual mobility map* because "the semantic information regarding collective behaviours of resumes can be easily perceived in such a compact view". In addition, he believed that resume validation is valuable and practical for real-world RA task; indeed, "there is no similar function existed in current RA tools".

Lastly, the experts pointed out that our system still has the following limitations which should be improved in the future.

1. The information on spatial dimension has not considered yet. Intuitively, the spatial career trajectory and spatial mobility on geographical map are also important.

2. The ego-spiral graph should be extended into a social network for better visual comprehension, because the relations among the surrounding nodes are also important.

3. Case study qualitatively evaluates the functionality of visualization phase of our system. However, the mining phase should be further validated via some quantitative experiments.

### 6. RELATED WORK
### 6.1 Resume Analysis

The prototype of resume analysis is personal experience analysis born during the World War II [4]. In 1996, resumes of scientific researchers were used to evaluate researchers' research ability and to discover technology transfer process [5]. Gaughan and Robin [6] compared career trends of scientists between different countries using resumes of 800 physicists and bio-scientists from France and the USA. Sabatier et al. [7] used resumes of 583 French bio-scientists to research what kind of impact career mobility had on various career progress patterns.

DaXtra [8] is a CV analysis software suite for recruitment management. DaXtra contains resume parsing, semantic searching and intelligent matching components which facilitate recruitment processes. Visualize.me [9] employs the infographic representation to turn a structured resume into several informative charts.

### 6.2 NLP and Data Mining

Much research has been conducted on NLP-related sub-tasks: Tokenization, Part-of-speech (POS) Tagging, Named Entity Recognition (NER), Sentiment Analysis, Word Segmentation etc. [10, 11, 12, 13, 14]. There are plenty of NLP toolkits, such as Stanford CoreNLP [10], UIMA [15], and GATE [16].

In data mining field, Classification and Association Rules Mining are related to our system since our system classifies people into various categories based on their experiences, and mines relations between people. Popular classification techniques include Decision Trees [17], Naïve Bayes [18] and Support Vector Machine [19]. Many algorithms for generating association rules were presented, such as Apriori [20], Eclat [21] and FP-growth [22]

### 6.3 Text Visualization

Multiple visualizations have been developed to represent information in text data [23, 24, 25, 26, 27, 28].

**Concrete text visualization.** Word cloud has been used to provide the content overview of corpus [29]. Cui et al. [23] proposed a hybrid tag cloud visualization which incorporated a trend chart to illustrate temporal content evolutions in a set of documents. Wattenberg and Viégas [30] proposed the Word Tree to represent information-retrieval results in a tree form in which each node denoted a keyword. Ham et al. [31] presented the Phrase Net to display a graph whose nodes are phrases and whose edges indicate that two phrases are linked by a specific relation.

**Abstract text visualization.** Havre et al. [32] proposed ThemeRiver to visualize thematic variations over time using a river-like metaphor. Luo et al. [33] used a similar metaphor to presents events based on event-based text analysis. Cui et al. [34] presented TextFlow for analyzing topical evolution patterns in which critical events were detected as flags of topic birth/disappearance and topic merging/splitting. Wu et al. [24] proposed OpinionFlow to visualize the diffusion of public opinions on social media by combining a Sankey graph with a density map.

### 7. CONCLUSION

This paper presents a visual analytics prototype based on resume data, aiming to help users understand the semantic patterns hidden in the resumes. First, a text-mining based approach is presented to extract the semantic information. Then, three visualization modules are designed to represent the semantic information in multiple perspectives. Through in-depth interaction and exploration performed by domain experts, our system are proved to be effective.

Further work can focus on the extension of the visualization on geographical dimension by considering the temporal-spatial mobility information. We also need to design better visual encodings for representing social relations and coordinated interaction between various views. Lastly, we consider to make a formal evaluation of our system, especially on mining performance.

### 8. ACKNOWLEDGMENTS

This work is supported by National Basic Research Program of China (2013CB329305), Natural Science Foundation of China (61303164, 61402447, 61502466) and K. C. Wong Education Foundation.

**Table 2. Scoring results for each task**

| Task# | 1 | 2 | 3 | 4 | 5 | 6 | 7 |
|---|---|---|---|---|---|---|---|
| **Score** (expert#1 / #2) | 4/3 | 4/5 | 4/4 | 5/5 | 5/4 | 5/5 | 5/4 |